\documentclass[twocolumn]{aastex7}

\newcommand{\kms}{km\,s$^{-1}$}
\newcommand{\msun}{\mbox{M$_{\odot}$}}

\newcommand{\Ni}{$^{56}$Ni}

\shorttitle{SN\,2023taz}
\shortauthors{A. Aamer et al.}
\usepackage{subcaption} 
\usepackage{amsmath}

\begin{document}

\title{SN 2023taz: Implications for the UV Diversity of Superluminous Supernovae}




\author[orcid=0000-0002-9085-8187,gname=Aysha, sname=Aamer]{Aysha Aamer} 
\affiliation{Astrophysics Research Centre, School of Mathematics and Physics, Queen’s University Belfast, UK}
\affiliation{Department of Astronomy and Steward Observatory, University of Arizona, 933 North Cherry Avenue, Tucson, AZ 85721-0065, USA}
\email[show]{aaamer01@qub.ac.uk}

\author[orcid=0000-0002-2555-3192,gname=Matt,sname=Nicholl]{Matt Nicholl}
\affiliation{Astrophysics Research Centre, School of Mathematics and Physics, Queen’s University Belfast, UK}
\email{}

\author[gname=Charlotte,sname=Angus]{Charlotte Angus }
\affiliation{Astrophysics Research Centre, School of Mathematics and Physics, Queen’s University Belfast, UK}
\email{}

\author[gname=Shubham,sname=Srivastav]{Shubham Srivastav}
\affiliation{Astrophysics, Department of Physics, University of Oxford, Keble Road, Oxford, OX1 3RH, UK}
\email{}

\author[gname=Jeff,sname=Cooke]{Jeff Cooke}
\affiliation{Centre for Astrophysics and Supercomputing, Swinburne University of Technology, Hawthorn, VIC, 3122, Australia}
\affiliation{ARC Centre of Excellence for Gravitational Wave Discovery (OzGrav), Victoria 3122, Australia}
\email{}

\author[gname=Natasha,sname=Van Bemmel]{Natasha Van Bemmel}
\affiliation{Centre for Astrophysics and Supercomputing, Swinburne University of Technology, Hawthorn, VIC, 3122, Australia}
\affiliation{ARC Centre of Excellence for Gravitational Wave Discovery (OzGrav), Victoria 3122, Australia}
\email{}

\author[gname=Fr\'{e}d\'{e}rick,sname=Poidevin]{Fr\'{e}d\'{e}rick Poidevin}
\affiliation{Instituto de Astrof\'{\i}sica de Canarias, V\'{\i}a L\'actea, 38205 La Laguna, Tenerife, Spain}
\affiliation{Universidad de La Laguna, Departamento de Astrof\'{\i}sica, 38206 La Laguna, Tenerife, Spain}
\email{}

\author[orcid=0000-0003-3154-2120,gname=Stefan,sname=Geier]{Stefan Geier}
\affiliation{GRANTECAN: Cuesta de San Jos\'{e} s/n, 38712 Bre{\~n}a Baja, La Palma, Spain}
\affiliation{Instituto de Astrof\'{\i}sica de Canarias, V\'{\i}a L\'actea, 38205 La Laguna, Tenerife, Spain}
\email{}

\author[orcid=0000-0003-0227-3451, gname=Joseph,sname=Anderson]{Joseph P. Anderson}
\affiliation{European Southern Observatory, Alonso de C\'ordova 3107, Casilla 19, Santiago, Chile}
\affiliation{Millennium Institute of Astrophysics MAS, Nuncio Monsenor Sotero Sanz 100, Off. 104, Providencia, Santiago, Chile}
\email{}

\author[gname=Thomas,sname=de Boer]{Thomas de Boer}
\affiliation{Institute for Astronomy, University of Hawai`i, 2680 Woodlawn
Drive, Honolulu HI 96822, USA}
\email{}

\author[gname=Kenneth,sname=Chambers]{Kenneth C. Chambers}
\affiliation{Institute for Astronomy, University of Hawai`i, 2680 Woodlawn
Drive, Honolulu HI 96822, USA}
\email{}

\author[orcid=0000-0002-1066-6098,gname=Ting-Wan,sname=Chen]{Ting-Wan Chen}
\affiliation{Graduate Institute of Astronomy, National Central University,
300 Jhongda Road, 32001 Jhongli, Taiwan}
\email{}

\author[orcid=0000-0002-1650-1518,gname=Mariusz,sname=Gromadzki]{Mariusz Gromadzki}
\affiliation{Astronomical Observatory, University of Warsaw, Al. Ujazdowskie 4, 00-478 Warszawa, Poland}
\email{}

\author[orcid=0000-0003-2375-2064,gname=Claudia,sname=Guti\'errez]{Claudia P. Guti\'errez}
\affiliation{Institut d'Estudis Espacials de Catalunya (IEEC), Edifici RDIT, Campus UPC, 08860 Castelldefels (Barcelona), Spain}
\affiliation{Institute of Space Sciences (ICE, CSIC), Campus UAB, Carrer de Can Magrans, s/n, E-08193 Barcelona, Spain}
\email{}

\author[gname=Erkki,sname=Kankare]{Erkki Kankare}
\affiliation{Department of Physics and Astronomy, University
of Turku, 20014 Turku, Finland}
\email{}

\author[0000-0002-8770-6764,gname=R\'eka,sname=K\"onyves-T\'oth]{R\'eka K\"onyves-T\'oth}
\affiliation{HUN-REN Research Centre for Astronomy and Earth Sciences, Konkoly Observatory, MTA Centre of Excellence, Konkoly Thege Miklós út 15-17., H-1121 Budapest, Hungary}
\affiliation{Department of Experimental Physics, Institute of Physics, University of Szeged, D\'om t\'er 9, Szeged, 6720 Hungary}
\email{}

\author[orcid=0000-0002-7272-5129,gname=Chien-Cheng,sname=Lin]{Chien-Cheng Lin}
\affiliation{Institute for Astronomy, University of Hawai`i, 2680 Woodlawn
Drive, Honolulu HI 96822, USA}
\email{}

\author[gname=Thomas,sname=Lowe]{Thomas B. Lowe}
\affiliation{Institute for Astronomy, University of Hawai`i, 2680 Woodlawn
Drive, Honolulu HI 96822, USA}
\email{}

\author[gname=Eugene,sname=Magnier]{Eugene Magnier}
\affiliation{Institute for Astronomy, University of Hawai`i, 2680 Woodlawn
Drive, Honolulu HI 96822, USA}
\email{}

\author[gname=Paolo,sname=Mazzali]{Paolo Mazzali}
\affiliation{Astrophysics Research Institute, Liverpool John Moores University, 146 Brownlow Hill, Liverpool L3 5RF, UK}
\affiliation{Max-Planck-Institut fur Astrophysik, Karl-Schwarzschild Straße 1, 85748 Garching, Germany}
\email{}

\author[orcid=0000-0001-7186-105X,gname=Kyle,sname=Medler]{Kyle Medler}
\affiliation{Institute for Astronomy, University of Hawai`i, 2680 Woodlawn
Drive, Honolulu HI 96822, USA}
\email{}

\author[gname=Paloma,sname=Minguez]{Paloma Minguez}
\affiliation{Institute for Astronomy, University of Hawai`i, 2680 Woodlawn
Drive, Honolulu HI 96822, USA}
\email{}

\author[gname=Tom\'as,sname=M\"uller-Bravo]{Tom\'as E. M\"uller-Bravo}
\affiliation{School of Physics, Trinity College Dublin, The University of Dublin, Dublin 2, Ireland}
\affiliation{Instituto de Ciencias Exactas y Naturales (ICEN), Universidad Arturo Prat, Chile}
\email{}

\author[orcid=0009-0005-8379-3871,gname=Ben,sname=Warwick]{Ben Warwick}
\affiliation{University of Warwick, Coventry CV4 7AL, UK}
\email{}

\begin{abstract}

Superluminous supernovae (SLSNe) are some of the brightest explosions in the Universe representing the extremes of stellar deaths. At the upper end of their distribution is SN\,2023taz, one of the most luminous SLSNe discovered to date with a peak absolute magnitude of $M_{g,\rm{peak}}=-22.75 \pm 0.03$ and a lower limit for energy radiated of $E=2.9 \times 10^{51}$\,erg. Magnetar model fits reveal individual parameter values typical of the SLSN population, but the combination of a low $B$-field and ejecta mass with a short spin period places SN\,2023taz in a unusual region of parameter space, accounting for its extreme luminosity. The optical data around peak are consistent with a temperature of $\sim$17\,000\,K but SN\,2023taz shows a surprising deficit in the UV compared to other events in this temperature range. We find no indication of dust extinction that could plausibly explain the UV deficit. The lower level of UV flux is reminiscent of the absorption seen in lower-luminosity events like SN\,2017dwh, where Fe-group elements are responsible for the effect. However, in the case of SN\,2023taz, there is no evidence for a larger amount of Fe-group elements which could contribute to line blanketing. Comparing to SLSNe with well-observed UV spectra, an underlying temperature of $8000-9000$\,K would match the UV spectral slope, but is not consistent with the optical colour temperatures of these events. The most likely explanation is enhanced absorption by intermediate-mass elements, challenging previous findings that SLSNe exhibit similar UV absorption line equivalent widths. This highlights the need for expanded UV spectroscopic coverage of SLSNe, especially at early times, to build a framework for interpreting their diversity and to enable classification at higher redshifts where optical observations will exclusively probe rest-frame UV emission.

\end{abstract}


\keywords{\uat{Supernovae}{1668} -- \uat{Core-collapse supernovae}{304} -- \uat{Massive stars}{732}}

\section{Introduction} \label{sec:intro}

Superluminous supernovae (SLSNe) are a subset of extreme stellar explosions with luminosities that far exceed those of ordinary stripped envelope SNe (SESNe). The bolometric peaks for the most luminous of these explosions easily exceed $10^{44}$\,erg\,s$^{-1}$ \citep{Gal-Yam2009, Quimby2011}. This is up to 100 times greater than the peak reached by typical SNe. 

These events cannot be explained by the radioactive decay of $^{56}$Ni alone, which is the main power source invoked to explain other SESNe. For SLSNe a $^{56}$Ni-powered model would require Ni masses comparable to or greater than the total ejecta mass \citep[e.g.][]{Inserra2013b}. Instead the leading theories propose interaction with circumstellar material (CSM), or a magnetar central engine. Although the former is used to explain hydrogen-rich SLSNe (SLSN-II) \citep{Smith2007b, Drake2010}, both have been proposed for hydrogen-poor SLSNe with the magnetar spin-down mechanism being widely adopted \citep[e.g.][]{Inserra2013b, Nicholl2015a, Gomez2024}. Both mechanisms can reproduce the broad light curves and high energies observed, but they imply very different progenitor systems and explosion physics. Disentangling these scenarios is critical for understanding SLSNe, probing the lives and deaths of the most massive stars, and the physical conditions of early star-forming environments.


The spectral energy distributions (SEDs) of SLSNe typically peak in the rest-frame $u$ or near-ultraviolet (UV) bands, indicative of high temperatures \citep[e.g.][]{Gomez2024}. However, in the UV, there are broad absorption lines from elements such as Mg, C, Si, and Fe which dominate the spectra \citep{Quimby2011, Chomiuk2011, Howell2013, Vreeswijk2014, Yan2018}. These features are thought to be formed in the outer layers on the ejecta due to their high velocities (typically $>10\,000$\,\kms). Previous studies have shown that the UV spectra of SLSNe tend to exhibit similar equivalent widths (EWs) in their absorption lines, regardless of their luminosity \citep{Nicholl2017d, Yan2017b}. The studies suggest that differing amounts of UV emission can generally be explained by varying power levels of a central energy source, rather than differences in absorbing column depths. This observation is consistent with a scenario where the continuum forms beneath the absorbing layer of fast moving material, rather than in an external collision with CSM \citep{Nicholl2017d}.

At high redshifts ($z \gtrsim 1$), the rest-frame UV part of a supernova (SN) spectrum is redshifted into the observer-frame optical. However, at these distances, the characteristic O II absorption features that are typically used to classify SLSNe are shifted to longer wavelengths where observations become increasingly challenging for ground-based telescopes \citep[e.g.][]{Barbary2009, Cooke2012, Pan2017, Smith2018, Curtin2019}. As a result, the detection and classification of SLSNe at high-$z$ must rely primarily on their rest-frame UV properties. Early discoveries revealed a surprisingly uniform class of SLSNe spectra in the near-UV, particularly between 2000–3000\,\AA, suggesting that UV diagnostics could provide a reliable classification framework across a wide range of redshifts \citep{Smith2018}. With the advent of deeper, wide-field surveys, the number of SLSNe detected at high-$z$ is expected to increase dramatically \citep{Villar2018}. This growth is not solely due to the larger volumes being probed, but also because SLSNe may be more common in the early universe, where the stellar mass distribution may be skewed towards more massive progenitors \citep{Cooke2012}. This underpins the urgent need to build a detailed understanding of the connection between UV and optical properties in low-$z$ SLSNe, as this will be essential for interpreting the large samples of high-$z$ SLSNe expected from upcoming surveys.

In this paper we present SN\,2023taz, one of the brightest SLSNe to date at optical wavelengths, but with a deficit in the near-UV compared to other SLSNe with similar colour temperatures in the optical. This challenges our current understanding of UV diversity in SLSNe and raises important questions about the central engine model, as well as our ability to identify SLSNe at high redshifts. The paper is structured as follows. Section \ref{sec:observations} presents the data collected for this object. The host galaxy for this event is analysed in Section \ref{sec:host}. Section \ref{sec:lc} discusses the analysis of the light curve, and Section \ref{sec:params} looks at the modelling of this light curve and the inferred parameters of the system. Section \ref{sec:spec_ev} discusses the analysis of the spectra. We discuss the possible scenarios for this object in Section \ref{sec:UV}. Lastly we present our conclusions in Section \ref{sec:conclusion}.

\begin{table}[t]
	\centering
	\caption[Spectroscopic observations of SN\,2023taz]{Spectroscopic observations of SN\,2023taz. Phase is given in rest frame frame days with respect to the time of maximum light in the $o-$band. Observations are obtained either on the NTT telescope using the EFOSC2 instrument, or on the NOT with the ALFOSC instrument.}
	\label{tab:spectra}
	\begin{tabular}{cccccc} 
		\hline \hline
            Date & MJD & Phase & Grism & Exposure & Telescope \\
              &  & (days) &  &  time (s) &  \\
            \hline
		04-11-2023 & 60252.3 & 8.0 &  Gr\#13 & 900 & NTT \\
		06-11-2023 & 60254.0 & 9.3 &  Gr\#16 & 2700 & NTT\\
            06-11-2023 & 60254.1 & 9.3 &  Gr\#11 & 2700 & NTT \\
            24-11-2023 & 60272.1 & 22.1 &  Gr\#13 & 2700 & NTT \\
            09-12-2023 & 60287.1 & 32.8 &  Gr\#13 & 2700 & NTT \\
            11-12-2023 & 60289.9 & 34.8 &  \#4 & 3x1000 & NOT\\
            18-12-2023 & 60296.1 & 39.2 &  Gr\#13 & 2700 & NTT  \\
            18-12-2023 & 60296.2 & 39.2 &  Gr\#13 & 2700 & NTT  \\
            16-01-2024 & 60325.1 & 59.7 &  Gr\#13 & 2700 & NTT \\
            16-01-2024 & 60325.1 & 59.8 &  Gr\#13 & 2700 & NTT  \\
            30-06-2024 & 60491.4 & 177.9 &  Gr\#13 & 2700 & NTT  \\
            30-06-2024 & 60491.4 & 178.0 &  Gr\#13 & 2700 & NTT \\
            30-06-2024 & 60491.4 & 178.0 &  Gr\#13 & 2700 & NTT \\
            06-07-2024 & 60497.3 & 182.2 &  Gr\#16 & 2700 & NTT \\
            06-07-2024 & 60497.4 & 182.2 &  Gr\#16 & 2700 & NTT \\
            02-01-2025 & 60678.1 & 310.6 & - & 2600 & VLT \\
            
		\hline \hline
	\end{tabular}
\end{table}

\section{Observations} \label{sec:observations}
\subsection{Discovery and Classification} \label{subsec:discovery}

SN\,2023taz was first detected by the Panoramic Survey Telescope and Rapid Response System \cite[Pan-STARRS;][]{Chambers2016} on 2023-09-12 under the internal name PS23ila at a magnitude $m_{w} = 20.36$\,mag. This followed a deep upper limit three days prior at $m_{w} \gtrsim 21.8$\,mag. It was later reported by the Asteroid Terrestrial-impact Last Alert System (ATLAS) project \citep{Tonry2018} on 2023-10-06 under the name ATLAS23tqv with a magnitude of $m_{c} = 19.14$\,mag, and by the Zwicky Transient Facility (ZTF) \cite[ZTF;][]{Bellm2019} on 2023-12-06 under the name ZTF23abgzmfs with a magnitude of $m_{g} = 20.03$\,mag. The event has a right ascension, declination of (J200) 19$^h$28$^m$23.01$^s$, -04$^\circ$54'43.86".

The object was chosen for classification by the Advanced Public ESO Spectroscopic Survey of Transient Objects \citep[ePESSTO+;][]{Smartt2015} based on the lack of visible host and long light curve rise. A spectrum obtained with the ESO New Technology Telescope on 2023-11-04 was consistent with a young SLSN \citep{Aamer2023}. A redshift of $z=0.42$ was determined using the SuperNova IDentification code \cite[\textsc{SNID};][]{Blondin2007}. This is a spectral cross-correlation tool used to classify supernovae by comparing observed spectra against a library of template spectra across various types and phases. Re-examination of the spectrum shows narrow [O III] $\lambda\lambda$4959, 5007 and H$\beta$ host galaxy lines at a redshift $z=0.407$. We therefore adopt this latter value for the analysis in this paper. A flat $\Lambda$CDM cosmology is also assumed throughout with $H_{0}=70$\,\kms\,Mpc$^{-1}$ and $\Omega_{\Lambda}=0.7$. This gives a luminosity distance of $D_{L}=2239$\,Mpc.

\subsection{Photometry} \label{subsec:photometry}

Photometry for SN\,2023taz was obtained from a number of telescopes. Data in $c$ and $o$ bands were collected from the ATLAS forced photometry server \citep{Tonry2018, Smith2020, Shingles2021}, and in $g$ and $r$ bands from the Lasair ZTF forced photometry server \citep{Masci2019}, both of which are performed on difference images. For the ATLAS data, observations were binned to a daily cadence. Pan-STARRS photometry in $w$ and $i$ bands was derived using the Pan-STARRS Image Processing Pipeline \citep{Magnier2020}.


Follow-up observations were also coordinated in \textit{ugriz} with the Las Cumbres Observatory (LCOgt). Within their network, several 1m telescopes were used across multiple observatories. These images were reduced using the BANZAI\footnote{https://github.com/LCOGT/banzai} pipeline. Late time photometry was obtained using the ESO New Technology Telescope (NTT) in \textit{gri} with the ESO Faint Object Spectrograph and Camera (EFOSC2). This was done through the ePESSTO+ collaboration. 

All photometry on raw frames for LCOgt and NTT data was carried out using the \textsc{photometry-sans-frustration} python package which performs point-spread function photometry using Astropy and Photutils \citep{Nicholl2023}. The frames are calibrated using the Pan-STARRS catalogue \citep{Flewelling2020} in order to obtain zeropoints. For the $u-$band images, zeropoints were calculated using the SkyMapper catalogue \citep{Onken2024}, with aperture photometry performed on reference stars within a cone of 10' from the target. Although the late time NTT observations resulted in 3$\sigma$ detections, the measured magnitudes are mainly consistent with marginal detections of (or upper limits on) the host galaxy in pre-explosion Pan-STARRS imaging (see Section \ref{sec:host}).

The Neil Gehrels Swift Observatory (\citealp[\textit{Swift};][]{Gehrels2004}) was also triggered to observe SN\,2023taz beginning on 2023-11-10 using the UV Optical Telescope \citep[UVOT;][]{Roming2005}, in the UV filters. However the SN was only detected in one epoch, in one filter with the rest providing upper limits. The individual frames in each filter were co-added using \texttt{uvotimsum} within the \textsc{HEASOFT} package. Photometry was performed on the final frames using the \texttt{uvotsource} task with a 5" aperture, following the procedure outlined in \citet{Poole2008} and \citet{Brown2009}. The measured count rates were converted to AB magnitudes using the UVOT photometric zero points \citep{Poole2008, Breeveld2011}.

Late-time imaging of the putative host galaxy was obtained using the Low Resolution Imaging Spectrometer \citep[LRIS;][]{Oke95,Steidel04} on 2025-08-31. LRIS is a two-arm instrument and the field was imaged with 3$\times$60\,s integrations (180\,s total exposure time per filter) in the $g-$band using the blue arm and the $i-$band using the red arm simultaneously by employing the D560 dichroic. Twilight sky images were taken for flat fielding. However, `supersky' flat field images for both filters were preferred and created by median-combining dithered images from two different fields using $g-$ and $i-$band filters separately. The images were aligned, stacked, and processed using conventional IRAF routines. Photometry was performed using Source Extractor \citep{1996A&AS..117..393B}, and zero point corrected by crossmatching nearby stars to measured values in the DECam Local Volume Exploration \citep[DELVE;][]{Drlica-Wagner:2021} Survey catalog.


\subsection{Spectroscopy} \label{subsec:spectroscopy}

Spectra for SN\,2023taz were obtained through ePESSTO+ using the NTT and EFOSC2 spectrograph. Spectra were taken using the Gr\#13 grism which observes between 3685-9315\,\AA. However, the data from 2023-11-06 were taken using both the Gr\#11 and the Gr\#16 grisms which range from 3380-7520\,\AA\ and 6015-10320\,\AA\, respectively. The spectra were first reduced using the PESSTO pipeline \citep{Smartt2015} which applies debiasing, flat-field correction, trace extraction, wavelength calibration, and flux calibration using standard stars observed with identical setups. A spectrum was also obtained on the Nordic Optical Telescope (NOT) on 2024-12-11 with the Alhambra Faint Object Spectrograph and Camera (ALFOSC). The observation consisted of 3 integrations of 1000 seconds using the Grism\#4 and a slit aperture of 1.3". A standard reduction and calibration of the spectrum was conducted using Pypeit \citep{Prochaska2020}.  

All spectra were then processed to remove tellurics using the python package \textsc{tellurics-begone}\footnote{https://github.com/EJRidley/tellurics-begone}. They were then corrected for foreground extinction using the \textsc{astropy} \citep{AstropyCollaboration2018} implementation of the \citet{Gordon2023} extinction law with $E(B-V)=0.0152$. The spectra were also flux corrected by scaling linearly to the closest photometry in the $g,r,$ and $i$ bands. Lastly, the spectra were redshift-corrected. For visual clarity, the [O III] $\lambda\lambda$ 4959,5007 host emission lines were removed by fitting a Gaussian centred on each wavelength within a 20\,\AA\ range, and subtracting them off. Spectra taken on the same night have been coadded, and all spectroscopic observations are listed in Table \ref{tab:spectra}. 

A late time spectrum was obtained on 2025-01-02 using X-Shooter on the Very Large Telescope (VLT). The UV, visible, and IR arms had total exposure times of 2440\,s, 2560\,s, and 2600\,s respectively. The data were then processed using the \texttt{soxspipe} package \citep{Young_soxspipe}, and corrected for MW extinction. We recover a continuum at signal-to-noise (SNR) $<1$ between $5000-6000$\,\AA, and cannot detect any clear transient features within the spectrum.                            

Extinction from the host galaxy was assumed to be negligible and has not been corrected for. This is based on the observed Balmer decrement (see Section \ref{sec:host}), as well as the observed colours and systematic studies of SLSN host galaxies, which show negligible or no host extinction \citep[e.g.][]{Perley2016, Schulze2018}. 

\section{Host Galaxy} \label{sec:host}


A faint galaxy designated PSO J030.8220-23.6798 is reported in the PS1 stacked object catalog in all of the $g,r,i,z,y$ bands, with a mean offset of $0.54"$ from the position of SN\,2023taz. We assume that this is the host galaxy. However the detection is only significant at the 3$\sigma$ level in $r$ band, with a PSF magnitude of $m_{r,\rm{host}}=22.90 \pm 0.19$\,mag and a Kron magnitude of $m_{r,\rm{host}}=22.99 \pm 0.26$\,mag. The measurement in $g$ is just below the threshold with a PSF magnitude of $m_{g,\rm{host}}=23.99 \pm 0.32$\,mag, and no corresponding Kron magnitude. The measured magnitudes in $i$ and $z$ are less certain with PSF magnitudes of $m_{i,\rm{host}}=23.85 \pm 0.52$\,mag and $m_{z,\rm{host}}=23.23 \pm 0.48$\,mag. Visually inspecting the images using the PS1 image cutout server shows no clear source at the position of the transient in these filters, confirming the marginal nature of these detections. Instead the LRIS images obtained at a phase of +481 days post peak were used to measure the host galaxy magnitude at $m_{i,\rm{host}}=23.20\pm0.08$ mag and $m_{g,\rm{host}}=23.97\pm0.09$ mag in $i-$ and $g-$bands respectively. The $r-$band photometry of SN\,2023taz has been host-subtracted using the Pan-STARRS flux measurement, and the $g-$ and $i-$band have been host-subtracted using the LRIS flux measurements. For the early phases of the light curve (before $\sim$100 days), we also apply host subtraction in the $z$ band. As the host detection is marginal, it does not make a significant difference to the light curve, as the SN flux dominates at these phases.

A galaxy K-correction calculator was used to calculate the absolute magnitudes of the host \citep{Chilingarian2010, Chilingarian2012}. Using this we estimate absolute magnitudes of $M_{g,\rm{host}}=-18.40 \pm 0.32$\,mag and $M_{r,\rm{host}}=-18.92 \pm 0.19$\,mag. We can compare these to typical SLSN host magnitudes from \citet{Schulze2018}, which show $M_{B} = -17.10 \pm 0.30$\,mag for hosts with $ z\lesssim 0.5$. 
The $g$-band host magnitude of SN\,2023taz is therefore at the brighter end of SLSN hosts in this redshift range. However, \citet{Schulze2018} also find that the host galaxies tend to be brighter at higher redshifts, so this difference may simply reflect the fact that SN\,2023taz lies at a higher redshift than the mean of the sample.
The host galaxy colour $g-r\approx0.5$ is consistent with other SLSN hosts at a similar redshift \citep{Lunnan2014}, suggesting it is a star-forming galaxy consistent with typical SLSN hosts.



\begin{figure*}
    \centering
    \includegraphics[width=2\columnwidth]{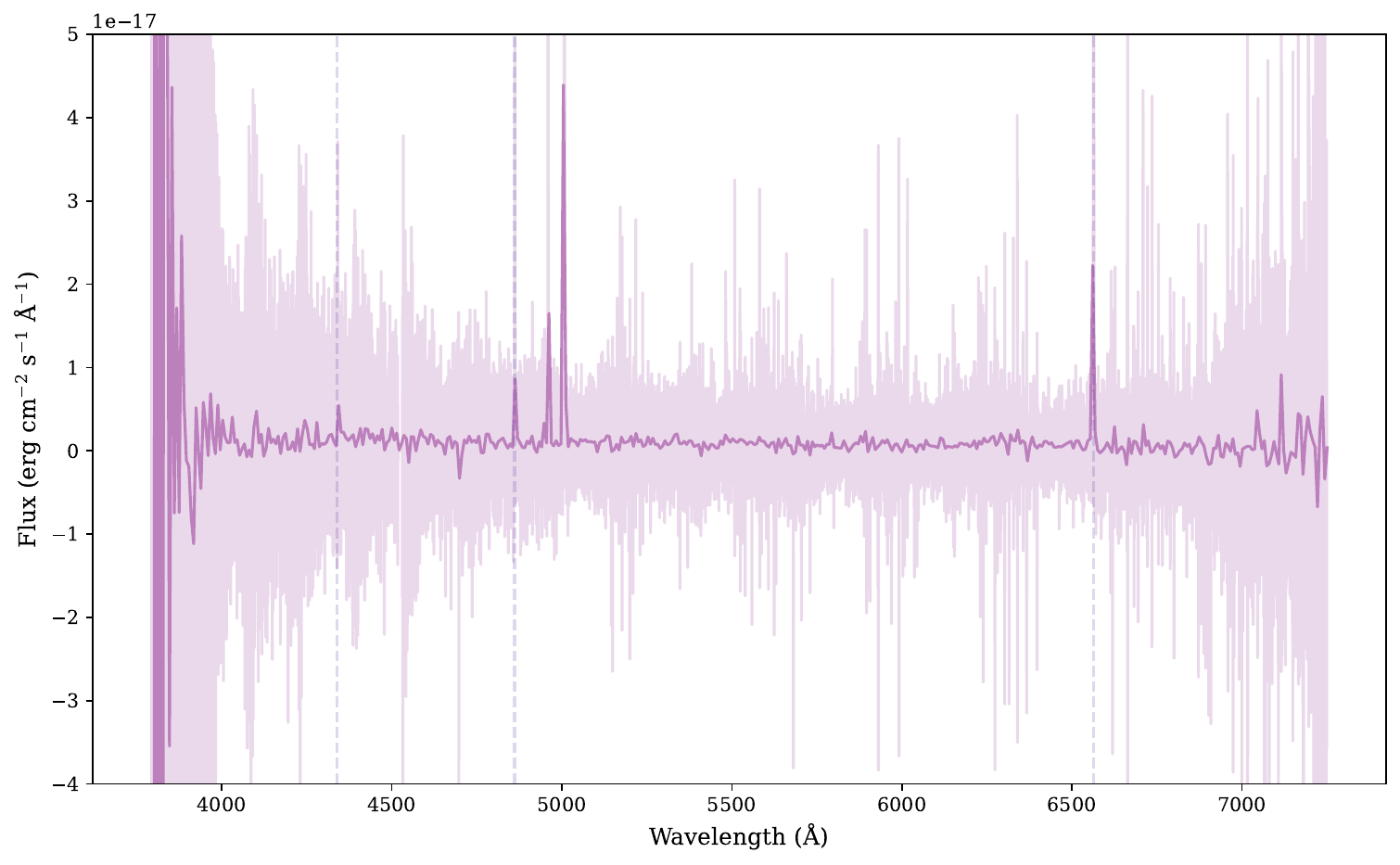}
    \caption{Spectrum of SN\,2023taz obtained with X-Shooter in the visible arm at a phase of +310.6 days post peak. Host galaxy emission lines from the Balmer series are highlighted with grey dashed lines.
    \label{fig:xshooter}}
\end{figure*}

A late time spectrum was obtained at a phase of +310.6 days using X-Shooter. The results from the visible arm are shown in Figure \ref{fig:xshooter} and clearly show the narrow host galaxy emission lines.  This spectrum shows very little SN signal except potentially a small bump $\sim$6300\,\AA\ which corresponds to a well observed nebular emission line of [O I] $\lambda\lambda$6300,6364. However, the signal is too weak for a definitive identification. The flux from the Balmer emission lines were measured by integrating over the emission profile. The continuum was determined by randomly sampling 100 points in the regions on either side of the line to obtain a median value for the continuum and its associated uncertainty. From this we find values of H$\alpha = (1.78 \pm 0.11) \times 10^{-16}$\,erg\,s$^{-1}$\,cm$^{-2}$, H$\beta = (5.88 \pm 0.45) \times 10^{-17}$\,erg\,s$^{-1}$\,cm$^{-2}$, and H$\gamma = (2.42 \pm 0.34) \times 10^{-17}$\,erg\,s$^{-1}$\,cm$^{-2}$. These yield Balmer decrements of H$\alpha/$H$\beta=3.03 \pm 0.29$, and H$\gamma/$H$\beta=0.41 \pm 0.07$. As the spectrum was corrected for MW extinction, any residual extinction present would be from the host and local environment of the SN. These values are consistent with the expected ratios for in a scenario with no extinction \citep{Osterbrock2006}.


The H$\alpha$ emission line can also be used as a diagnostic for the star formation rate (SFR) within a galaxy \citep{Kennicutt1998}. This yields a value of SFR = $0.85 \pm 0.05\,\rm{M}_{\odot}$\,yr${-1}$. The SFR is consistent with the sample of hosts from \citet{Leloudas2015b} where a range of $0.01 -6.04\,\rm{M}_{\odot}$\,yr${-1}$ was found.

\section{Light Curve} \label{sec:lc}

\begin{figure*}
    \centering
    \includegraphics[width=2\columnwidth]{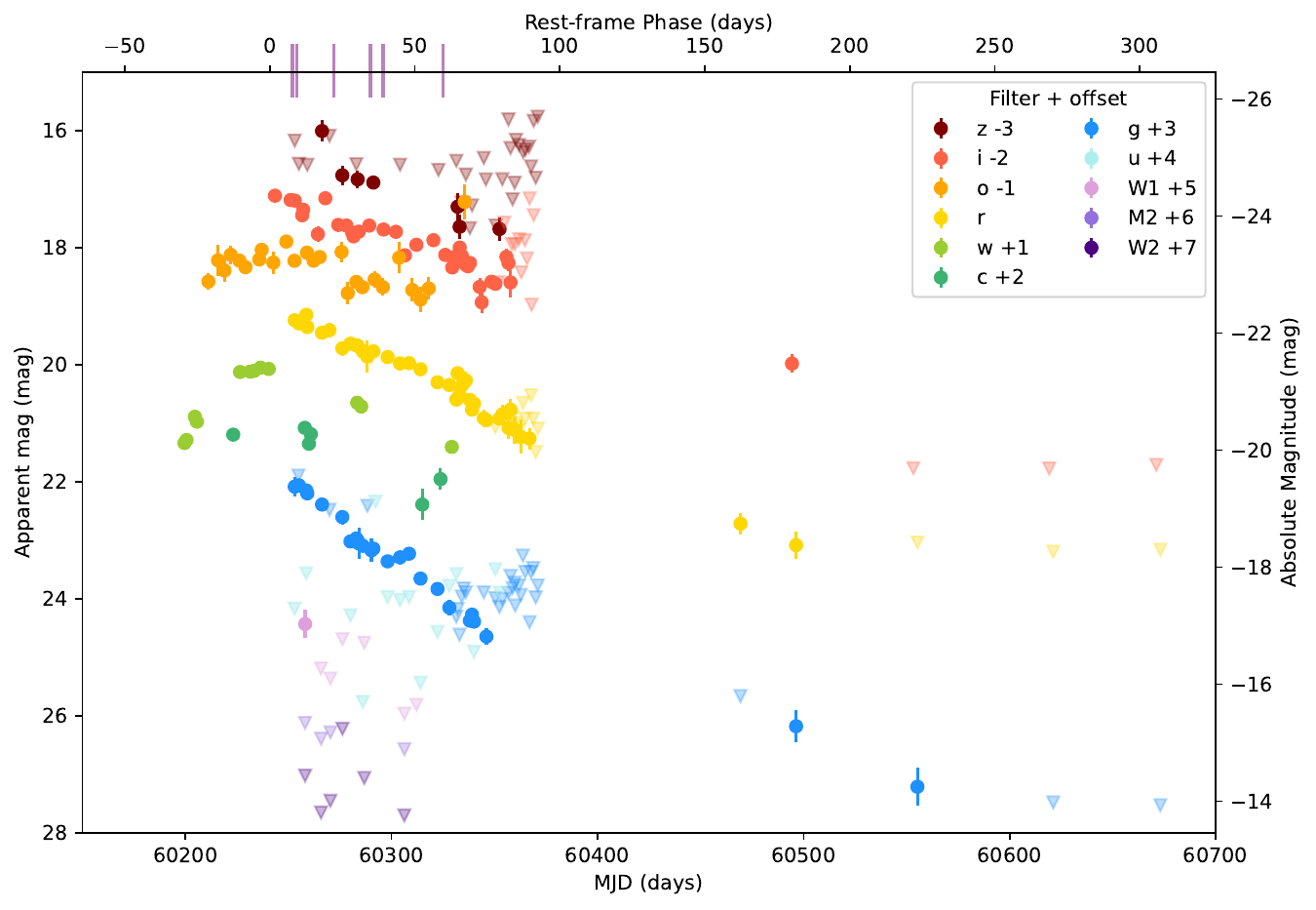}
    \caption{The light curve of SN\,2023taz, with all magnitudes in the AB system and uncorrected for Milky Way extinction. Phases are given in rest-frame days relative to the $w-$band maximum. The figure compiles photometric data from ATLAS, Pan-STARRS, ZTF, Swift, LCOgt, and NTT. MJD is in the observer frame, with 3$\sigma$ upper limits marked as inverted triangles. For clarity, ATLAS $c$ and $o$-band points are shown without upper limits and have been binned to a daily cadence. Vertical tick marks on the top axis indicate the epochs of the spectra displayed in Figure \ref{fig:spectra}.
    \label{fig:lc}}
\end{figure*}

The light curve for SN\,2023taz spanning over 200 days is presented in Figure \ref{fig:lc}.
The rise and peak of the light curve is detected at high SNR in the $w-$band data from Pan-STARRS. This is a broadband filter covering $g,r,$ and $i$ with an effective wavelength close to that of $r$ (or to rest-frame $g$-band at the redshift of SN\,2023taz). To find the time of maximum light, we fit a 2nd degree polynomial and found a time of peak at MJD 60241 at an apparent magnitude $w_{\rm{peak}}=19.07$\,mag. Using the first spectrum obtained on MJD 60252.3, we perform a K-correction and find an absolute magnitude of $M_{g,\rm{peak}}= -22.75 \pm 0.03$\,mag (also corrected for a foreground extinction $A_{V} = 0.039$; \citealt{Schlafly2011}). This is at the upper extreme of SLSNe luminosities. \citet{Chen2023b} find a median peak luminosity of $M_{g,\rm{med}}=-21.48^{+1.13}_{-0.61}$\,mag from their sample of 78 SLSNe from ZTF, making this event more than 2$\sigma$ brighter than the average ZTF SLSN. More recently \citet{Gomez2024} analysed a sample of 262 SLSNe, from which the median peak rest-frame $g-$band absolute magnitude is measured to be $M_{g,\rm{med}} = -21.5^{+1.0}_{-0.6}$\,mag, consistent with the ZTF sample. SN\,2023taz is once again outside of the 2$\sigma$ range and is in fact brighter in the $g-$band than all events within both samples.

Figure \ref{fig:lc} shows the bluer bands fading much faster than the redder bands after peak, as the SN cools. This can be quantified by the time taken for the SN to fade by a factor of e in each band $\tau_{e}$. This e-folding decline time is $\tau_{e,g}=44$\,days and $\tau_{e,o}=82$\,days in the $g-$ and $o-$ bands respectively. This is comparable to the average decline time found in \citet{Gomez2024} of $\tau_{e,r}=44^{+38}_{-18}$\,days. This measurement also assumes that the first $g-$band point in the light curve is close to or at peak. The rapid cooling of SN\,2023taz also resulted in mostly UV non-detections. 



\begin{figure}
    \includegraphics[width=\columnwidth]{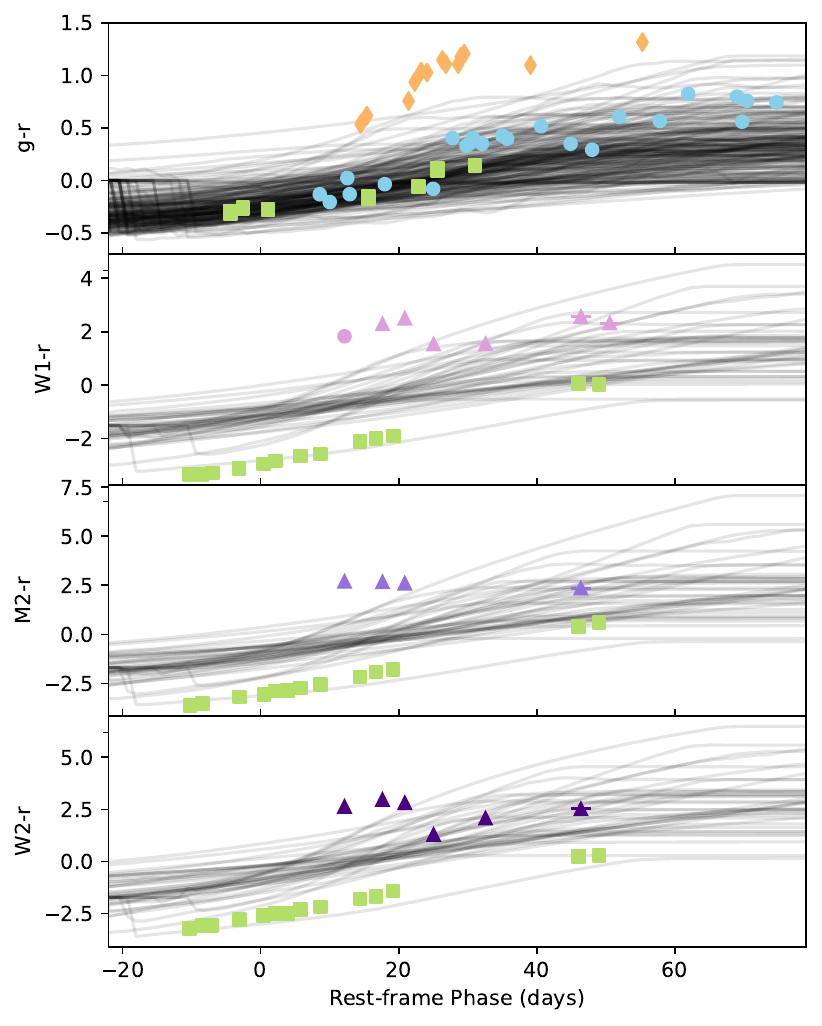}
    \caption{Photometric colours for SN\,2023taz in blue, pink, purple, and indigo compared to the colours from the sample in \citet{Gomez2024} shown in black. Measurements given in AB mag with lower limits denoted by triangles. Photometric colours for Gaia16apd \citep{Kangas2017, Yan2017b, Nicholl2017d} and SN\,2017dwh \citep{Blanchard2019} are also plotted in green squares and orange diamonds respectively. Phases are given relative to peak.
    \label{fig:colour}}
\end{figure}

\subsection{Colours} \label{sec:colour}

The colour evolution of SN\,2023taz is shown in Figure \ref{fig:colour}. This is compared to the model light curves of a subset of events from \citet{Gomez2024} with observations in each respective band. Within the sample, there are 44, 43, and 42 events with photometry in $W1, M2$, and $W2$, respectively. The photometry for SN\,2023taz was first K-corrected using the \textsc{slsne} python package \citep{Gomez2024} to ensure consistent colour comparisons between objects. The colours in the $g$ and UVOT bands with respect to $r$-band were then determined by subtracting the closest $r-$band magnitude. In all cases, the phase differences were less than 1.5 days. The $g-r$ colour starts off similar to the population average with $g-r \sim -0.3$\,mag at peak. It then evolves smoothly to redder colours before settling at $g-r \sim 0.8$\,mag. This late-time optical colour is somewhat redder than typical SLSNe. However, the UV-optical colours are the reddest in the population especially at early times. The triangles in the UV bands represent lower limits for the colour in each subplot. We can see this event is redder than average across the entire time span where UV photometry was obtained. It is especially apparent between $0-20$ days where the lower limits are constraining. The single $W1$ detection places tight constraints on the colour $W1-r=1.83 \pm 0.23$\,mag at 12 days post peak. At 12 days post peak the median and 1$\sigma$ spread is $W1-r = -0.73^{+0.39}_{-0.26}$\,mag for the subset of events with $W1$ photometry. This means the detection at that phase is more than 5$\sigma$ away from the median, pushing it to extremes of the population. 
Therefore, SN\,2023taz stands alone among the current sample as a highly luminous SLSN, with a very red UV-optical colour.

\begin{figure}
    \centering
    \includegraphics[width=\columnwidth]{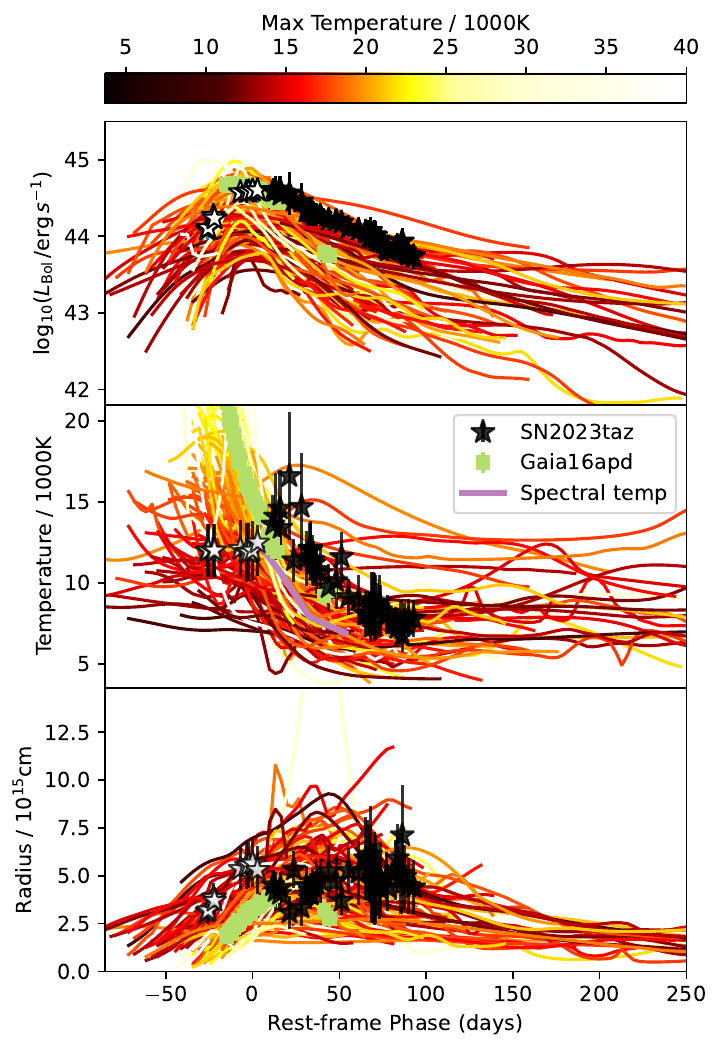}
    \caption{Parameters derived from SED fits of the light curve. Unfilled stars indicate where less than three bands were available for the fit. The colours of each line correspond to the peak blackbody temperature calculated for each event from the sample in \citet{Gomez2024}, and the parameters for Gaia16apd are plotted in green squares. 
    \textit{Top:} Bolometric luminosity. 
    \textit{Middle:} Blackbody temperature. The temperature from blackbody fits to the spectra are also plotted in purple. 
    \textit{Bottom: }Blackbody radius.
    \label{fig:bb_params}}
\end{figure}

In comparison Gaia16apd, an event with a comparable peak bolometric luminosity and optical light curve timescale, displays a much bluer colour than SN\,2023taz \citep{Kangas2017, Yan2017b, Nicholl2017d}. Its $g-r$ colour is initially comparable, but its UV colours are bluer by $\sim4$\,mag. This is not simply a consequence of Gaia16apd having a higher temperature. As shown in Figure \ref{fig:bb_params}, we find similar best-fitting blackbodies for the optical photometry of Gaia16apd and SN\,2023taz. Instead, the difference of several magnitudes in $UV-r$ colours seems to indicate a difference in their UV properties. The reddest SLSN in $g-r$ at late times is SN\,2017dwh \citep{Blanchard2019}. This event is similar to the general population at early times around peak, before diverging quickly and becoming much redder with a colour $g-r = 1$\,mag by 30 days post peak. There are no UV observations of this event, and therefore a comparison cannot be made for the UV colours. However, its spectrum shows evidence for strong Fe-group absorption in the near-UV \citep{Blanchard2019}; see Section \ref{sec:spec_ev}.

\subsection{Blackbody Parameters} \label{sec:BB_params}

The blackbody parameters for this analysis were calculated using the Python package \textsc{Superbol} \citep{Nicholl2018a}. This is a tool designed to derive bolometric light curves and fit blackbody models to the observed photometry. It works by constructing the SED at each epoch, integrating this to determine the optical luminosity, and fitting a blackbody curve to estimate the temperature, radius and unobserved UV/NIR luminosity of the emitting source. In this case, a modified blackbody (based on \citealp{Nicholl2017a} and \citealp{Yan2018}) was fit to the photometry. This suppresses the SED below a given wavelength following the form:

\begin{equation}
    f_{\lambda}(T, R) = 
    \begin{cases}
      (\frac{\lambda}{\lambda_{0}})^{\alpha} f_{\lambda, \text{BB}}(T, R) \text{\qquad \qquad \qquad for $\lambda < \lambda_{0}$} \\
      f_{\lambda, \text{BB}}(T, R) \text{\qquad \qquad \qquad \qquad \enspace for $\lambda > \lambda_{0}$}
    \end{cases}
    \label{eq:SED}
\end{equation}

Here $f_{\lambda}$ is the flux at a given wavelength $\lambda$, $\alpha$ is the suppression index, $\lambda_{0}$ is a cutoff wavelength for the suppression, and $f_{\lambda,\text{BB}}$ is the normal blackbody spectrum without any modifications applied. We use values of $\alpha=1.56$ and $\lambda_{0}=3072$\,\AA\ based on light curve modelling (Section \ref{sec:params}). The SEDs were also shifted into the rest-frame. We used the $w-$ and $r-$bands as reference points to anchor the colour evolution, allowing \textsc{Superbol} to infer the colours of the other bands either by assuming a constant colour evolution, or by interpolating unevenly-sampled photometry using a polynomial.

Integrating under the SED at each epoch also gives the bolometric luminosity at each point in time. By integrating this luminosity across the entire observed span on the SN, we estimate the total radiated energy is $E > 2.9 \times 10^{51}$\,erg. This value represents a lower limit, as it does not account for incomplete temporal and wavelength coverage, nor any emission occurring outside the span of our photometric observations. This energy is consistent with the median radiated energy of SLSNe $\sim1.3\times10^{51}$\,erg, with all known H-poor SLSNe having radiated $<5\times10^{51}$\,erg \citep{Gomez2024}. 

Figure \ref{fig:bb_params} shows the blackbody parameters in comparison to the full sample of events from \citet{Gomez2024}. We also highlight Gaia16apd on this plot. The unfilled stars represent the early light curve of SN\,2023taz, where there is only coverage in the $w,c,$ and $o$ bands. Since $w$ is effectively $c+o$, this means there is only a small part of the SED covered and therefore the blackbody fits are not reliable.

In the top panel of Figure \ref{fig:bb_params}, we can see that the bolometric luminosity for this event is at the upper extreme of the population. At peak it has a bolometric luminosity of $L_{\rm{bol}} = (4.1  \pm 1.7) \times 10^{44}$\,erg\,s$^{-1}$. At late times SN\,2023taz remains within the 1$\sigma$ range of the sample. It is also interesting to note that the brighter events tend to be the hottest, as evidenced by the colour scaling on the figure.


The middle panel of Figure \ref{fig:bb_params} shows the temperature evolution of the sample of events as estimated from these SED fits. SN\,2023taz reaches a peak temperature of $16\,600\pm4000$\,K, at 10 days after the bolometric peak. However hotter events at peak also cool quicker. This corresponds to the rapid spectral change between the ``hot" photospheric phase and the ``cool" photospheric phase as outlined in \citet{Quimby2018} and \citet{Aamer2025}. Plotted on the same subfigure in purple is the temperature derived from fitting a blackbody curve to the spectra (see Section \ref{sec:spec_ev}). In comparison, Gaia16apd starts off at a much hotter temperature of $\sim$22\,000\,K with a much more rapid decline to $\sim$15\,000\,K at peak. At this point the temperatures are comparable and shows that the UV-optical colour differences seen after peak in Section \ref{sec:colour} are not driven by different temperatures. The temperatures of both events are at the higher end but well within the 1$\sigma$ range of the population.

From the blackbody fits to the SED we can also calculate the size of the photospheric radius with time. The SLSN has a slow rise to peak radius, reaching a maximum radius of $R= (6.1 \pm 2.5) \times 10^{15}$\,cm around 60 days post peak. In comparison, Gaia16apd has a much smaller maximum radius which peaks soon after maximum light. However, at the peak of the light curve, both events have similar sized radii. As the temperature evolution of both events post peak is similar, this larger radius explains why the light curve of SN\,2023taz is much longer lived.


\section{Magnetar Parameters}\label{sec:params}

\begin{figure*}
      \centering
      \includegraphics[width=\linewidth]{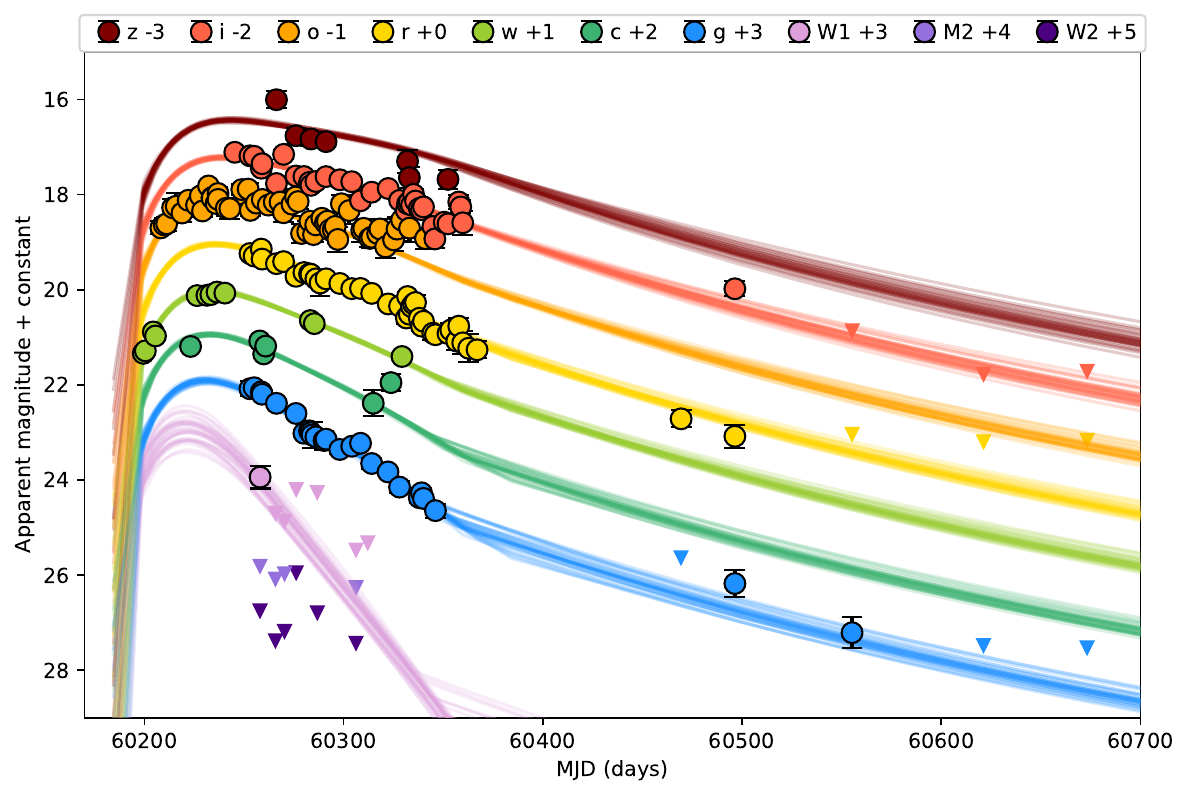}

    \caption{\textsc{MOSFiT} fits to the light curve of SN\,2023taz using the \texttt{slsnni} model. This model combines the magneter engine with a flexible level of contribution from the radioactive decay of \Ni. Upper limits are indicated via inverted triangles.
    \label{fig:mosfit_lcs}}
\end{figure*}

\begin{figure*}
      \centering
      \includegraphics[width=1\linewidth]{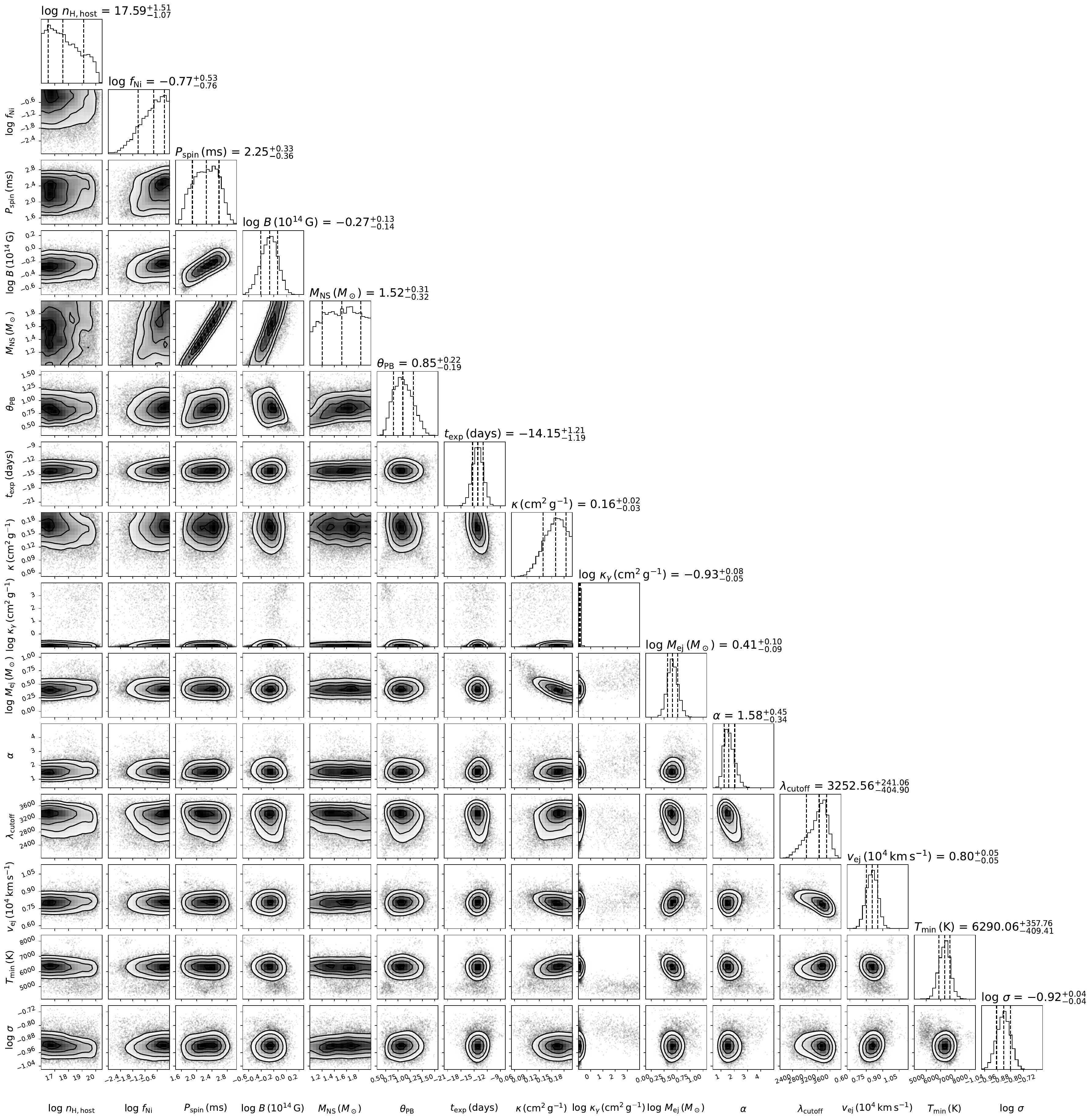}
    \caption{Corner plot of the posteriors for the \texttt{slsnni} model fit to the light curve of SN\,2023taz. Median and 1$\sigma$ ranges are labelled.
    \label{fig:mosfit_params}}
\end{figure*}


SLSNe show a wide range of luminosities and timescales. This requires a mechanism that decouples the magnitude of the heating source from its duration, and is largely independent of the ejecta properties. The simplest model that achieves this is the magnetar central engine model, which reproduces the bulk features of large SLSN samples \citep{Kasen2010, Mazzali2016, Nicholl2017a, Gomez2024} mainly through variations in spin period (heating rate), magnetic field (heating duration), and ejecta mass (diffusion timescale). As SN\,2023taz stands out as one of the brightest SLSNe to date, it is interesting to determine which features of the model would lead to such a bright (and relatively long-lived) event. While other power sources may be applicable to SLSNe, fitting SN\,2023taz within the same framework as other events enables more general insights, and fitting the multi-colour photometry allows us to test whether it is subject to anomalous extinction compared to other events.

To explore the central engine properties, we modelled the multi-colour light curve using the Modular Open Source Fitter for Transients (\textsc{mosfit}) \citep{Guillochon2018}. Specifically we used the \texttt{slsnni} model which models a magnetar internal engine following \citet{Nicholl2017a}, whilst allowing contribution from the radioactive decay of $^{56}$Ni as a free parameter. The model assumes a modified blackbody in the form of Equation \ref{eq:SED}, with $\lambda_0$ and $\alpha$ set as free parameters. This model has been widely applied to SLSNe \citep{Gomez2024} and shown to reproduce their overall light curve properties. For this work, we adopt the default set of priors, following the approach of \citet{Gomez2024}.


The best fit \textsc{mosfit} model light curve is shown in Figure \ref{fig:mosfit_lcs}. The rise and peak in the $o-$ and $w-$bands are well captured, as well as the decline in all bands. The exception to this is the $z-$band, where the model consistently overpredicts the luminosity, even at early times. This discrepancy may arise from an unconstrained host contribution that was not subtracted. Although it was assumed that the host flux would be negligible compared to the supernova at these early phases, this may have nevertheless affected the model fits. Although we show UVOT upper limits in the plot, we only include the single epoch with a detection in $UVW1$ in our fit, as the model struggles to fit the optical light curves and the deep UV limits simultaneously. The inability for the model to fit the data could indicate that the simple suppressed blackbody SED model does not fully account for strong line absorption in the UVOT $UVM2$ and $UVW2$ bands.

\begin{figure}
    \begin{subfigure}{1\columnwidth}
        \includegraphics[width=0.95\columnwidth]{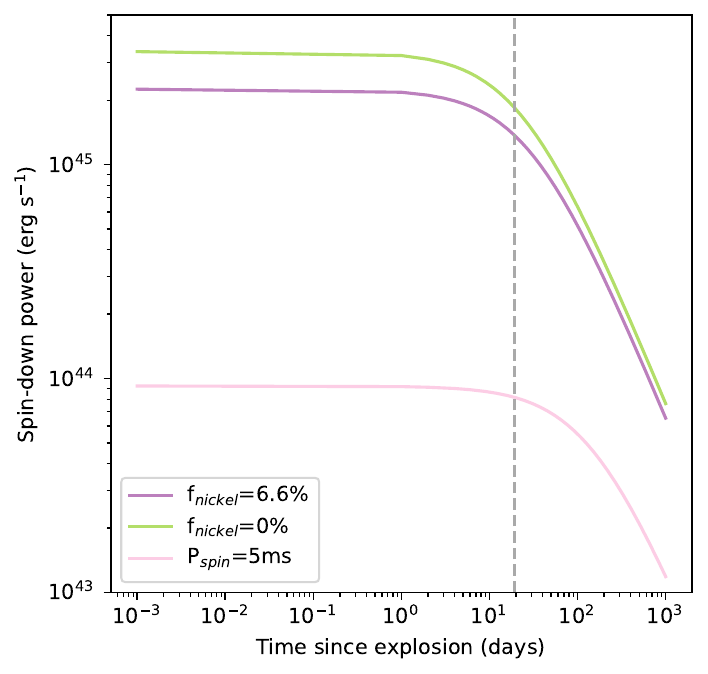}
        \caption{Magnetar spin-down energy based on the best fitting parameters when allowing the nickel fraction to vary in the models, as well as when the nickel fraction is fixed to 0\%. In pink is the spin down energy using the same parameters as the flexible nickel model but setting the spin period to a larger value of 5\,ms.}
        \label{fig:spin_down}
    \end{subfigure}
\bigskip
    \begin{subfigure}{\columnwidth}
        \includegraphics[width=0.95\columnwidth]{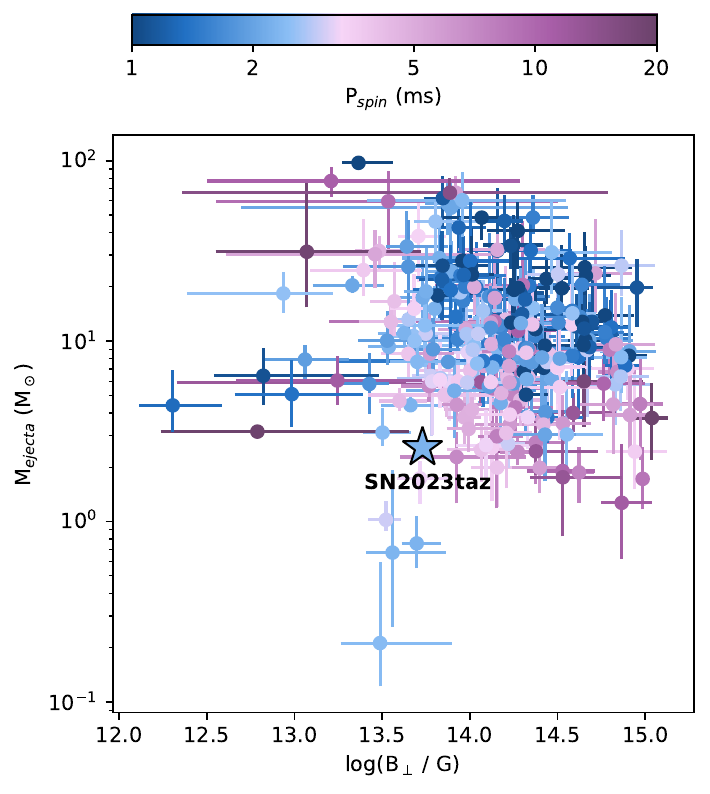}
        \caption{The perpendicular component of the magnetic field ($B_{\perp}$ against the ejecta mass ($M_{\rm{ej}}$) for the sample of SLSN from \citet{Gomez2024}. The points are also coloured by their spin period ($P_{\rm{spin}}$). The values derived for SN\,2023taz are plotted with a star shape.}
        \label{fig:B_vs_M}
    \end{subfigure}
\caption{Parameter space of the \textsc{MOSFiT} models.}
\label{fig:param_space}
\end{figure}

The best fit parameters for the \texttt{slsnni} model are shown in Figure \ref{fig:mosfit_params}. These parameters are mostly just outside the $1\sigma$ range of the population sample from \cite{Gomez2024}, including the spin period at $P_{\rm{spin}} = 2.25^{+0.33}_{-0.36}$\,ms just below the sample mean at $P_{\rm{spin, mean}} = 2.4^{+3.0}_{-1.2}$\,ms. A lower value of $P_{\rm{spin}}$ results in a more luminous event. The value for the perpendicular component of the magnetic field to the spin axis, $\log(B_{\perp}/\rm{G}) = 13.73^{+0.13}_{-0.14}$, is also just below the 1$\sigma$ range of values in \citet{Gomez2024} of $\log(B_{\perp, \rm{mean}}/\rm{G}) = 14.2 \pm 0.4$. The effect of a lower value for $B_{\perp}$ is a longer magnetar spin down timescale.

In our host galaxy and photospheric temperature analyses we assumed negligible extinction from the host galaxy. This is supported by the Balmer decrement in the in the host galaxy spectrum. It is also supported by the posterior of our light curve model fit, which independently fits the host extinction through the parameter for column density in the host (n$_{\rm{H,host}}$). The fits favoured n$_{\rm{H,host}} \leq 18$\,cm$^{-2}$, which corresponds to an extinction $A_V < 10^{-3}$\,mag \citep{Guver2009}.

The model also estimates the fraction of $^{56}$Ni ($f_{\rm{Ni}}$) in the ejecta required in addition to the contribution from the magnetar. This resulted in a value of $f_{\rm{Ni}} = 17.0^{+40.5}_{-14.0}\%$. We can combine this with the mass ejected parameter ($M_{\rm{ej}}=2.57^{+0.66}_{-0.48}$\,\msun) to find that the mass of \Ni\ ejected is $M_{\rm{Ni}}\sim$0.4\,\msun\ and could be as high as $M_{\rm{Ni}}\sim$1\,\msun based on the upper end of the uncertainty range. The fraction (and hence the mass) is not well constrained because the model prefers the magnetar as the primary heating source, and a \Ni\ mass up to $\sim$0.4\,\msun\ has little effect on the magnetar-powered light curve. To confirm this, we ran another fit with a fixed $f_{\rm{Ni}}=0\%$ which resulted in very similar fits and posteriors. This is consistent with the findings in \citet{Gomez2024} where most SLSNe do not require a significant contribution from radioactive decay to power them.

Figure \ref{fig:spin_down} shows the evolution of the heating rate, using the best fit spin period ($P_{\rm{spin}} = 2.25$\,ms), magnetic field ($B_{\perp} = 5.37 \times 10^{13}$\,G) and neutron star mass ($M_{\rm{NS}} = 1.52$\,\msun) from \textsc{MOSFiT}. The heating rate follows an initial plateau lasting $\sim$1.5\,days, followed by a steep decay with a power-law index of 2. The energy input is very similar whether or not a contribution from $^{56}$Ni is included, demonstrating that SN\,2023taz can easily be powered without a substantial contribution from radioactive decay. 

On the figure, we mark the time of bolometric peak with a vertical dashed line. The time of peak here is measured by averaging the bolometric light curves produced by 100 walkers from the \textsc{MOSFiT} output. The rise time measured for this event is $t_{\rm{rise}}=26$\,days, close to the median value found by \citet{Gomez2024} of $t_{\rm{rise,all}}=27^{+25}_{-13}$\,days. At peak luminosity, the observed radiated luminosity is comparable to the rate of energy input from the central engine \citep{Arnett1982}. Figure \ref{fig:spin_down} demonstrates that at the time of peak in SN\,2023taz, the energy injection of the magnetar remains high, and the spin-down power has not yet transitioned to the steep decline of the power-law tail. This is a consequence of the low $B_{\perp}$ value discussed earlier, which extends the spin-down timescale. The sustained energy injection up to the time of peak is what drives SN\,2023taz to be a very luminous SLSN \citep{Nicholl2017a}.

Figure \ref{fig:B_vs_M} shows the distribution of $B_{\perp}$ and $M_{\rm{ej}}$ values from \citet{Gomez2024}, with points coloured by their $P_{\rm{spin}}$ values. We can see that SN\,2023taz falls below the bulk of the population both in terms of ejecta mass, and B-field, lying on the edge of the joint distribution. The effect of a low $M_{\rm{ej}}$ is to reduce the rise time, and in this case matching it to the spin down timescale. However, other SLSNe occupy a similar region in $M_{\rm{ej}}-B_{\perp}$, and are generally less luminous than SN\,2023taz. The reason is that most of these events have longer spin periods $\gtrsim 4$\,ms, as shown by the colour scale. SN\,2023taz stands out by having a shorter spin period than most other objects in a similar region of the parameter space. This has a dramatic effect, as the magnetar peak spin-down rate scales as $P_{\rm{spin}}^{-4}$. We demonstrate the effect of spin period in Figure \ref{fig:spin_down} by plotting the spin down energy for a hypothetical event with the same $B_{\perp}$ as SN\,2023taz, but a longer spin period of $P_{\rm{spin}} = 5$\,ms. This results in a slightly longer spin-down time, but with a much lower level of energy throughout, resulting in a less luminous peak.

\section{Spectral Evolution}\label{sec:spec_ev}

\begin{figure*}
    \centering
    \includegraphics[width=2\columnwidth]{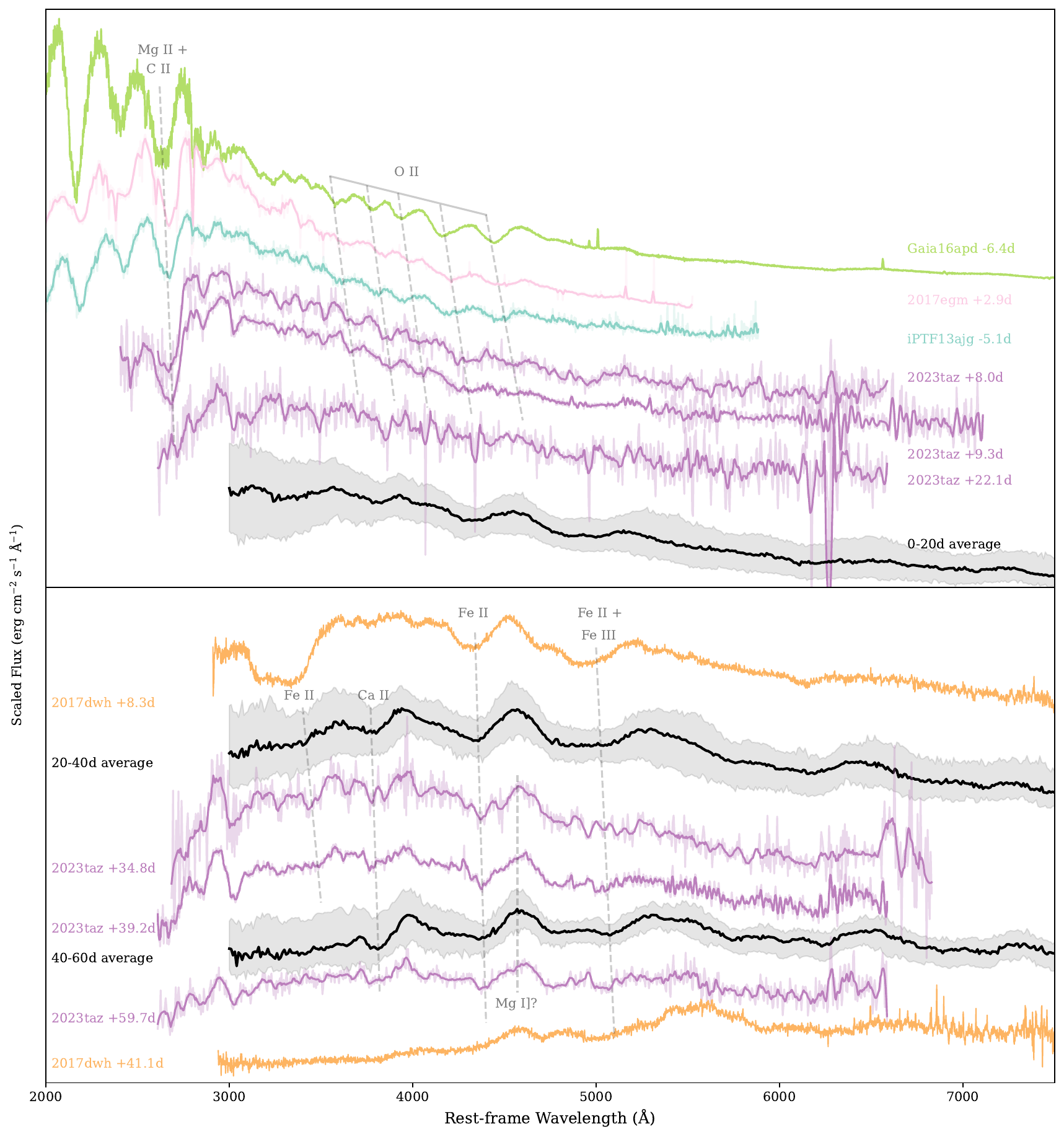}
    \caption{The spectral evolution of SN\,2023taz. Comparison spectra of Gaia16apd, SN\,2017egm, iPTF13ajg, and SN\,2017dwh are also plotted. Average spectra from \citet{Aamer2025} are plotted in black. Line identifications are highlighted with grey dashed lines.
    \label{fig:spectra}}
\end{figure*}

Figure \ref{fig:spectra} presents the post maximum spectroscopic evolution of SN\,2023taz. The spectra at early times are dominated by a hot, blue continuum, with relatively weak absorption features from O II superimposed. The spectral coverage extends to 53 days post peak, by which point the SN has transitioned from the hot photospheric phase to the cooler photospheric stage. The spectra still exhibit underlying continuum emission, but with O II replaced by O I, Mg I, Fe II and Ca II. These two distinct phases are split into the top and bottom panel of Figure \ref{fig:spectra} respectively. Unfortunately, the coverage does not extend to the nebular phase of this object.

Blackbody models were also fit to these spectra, and included in Figure \ref{fig:bb_params}. For the two spectra just after peak, the fits were trimmed to only include wavelengths above 2800\,\AA. This was done due to the deep UV absorption line just below this cut off, which interfered with the fits and forced a lower temperature fit that did not capture the shape in the optical. Applying a modified blackbody fit did not alter the temperatures significantly, with a largest deviation of $\sim700$\,K. These temperatures are slightly lower than those from the broadband SED fitting, but this may be due to the spectra not extending as far blue and therefore not capturing the peak of the SED well. This means we are only fitting to the Rayleigh-Jeans tail in these spectra, making the temperature much harder to constrain.

Examining the early spectra at 8-22\,days after peak, in the optical we can see the O II absorption lines blueshifted with a velocity of 6000\,km\,s$^{-1}$. This is lower than typical line velocities measured for other SLSNe typically around $10\,000-15\,000$\,\kms\ \citep[e.g.][]{Quimby2018, Gal-Yam2019a}. The lines are also quite weak compared to the continuum. The formation of the O II lines are found to be temperature dependent, with the classic ``W" shape appearing for temperatures in the range $14\,000-16\,000$\,K \citep{Konyves-Toth2022, Konyves-Toth2023, Saito2024}. From SED fitting in Section \ref{sec:BB_params} we can see that the temperature at the phases of these spectra are below this threshold, explaining the weaker O II lines.

We show line identifications in Figure \ref{fig:spectra} extending into the near UV. Here, we can see a broad absorption line with a minimum at 2672\,\AA. This line has been attributed to Mg II $\lambda\lambda$2798, 2803 \citep{Quimby2011, Chomiuk2011} with a possible contribution from C II \citep{Vreeswijk2014, Howell2013, Mazzali2016}.

Figure \ref{fig:spectra} also shows average SLSN spectra in different time bins, constructed by \citet{Aamer2025} based on a sample of 234 SLSNe. Comparisons have been plotted with phases corresponding to the phases of SN\,2023taz. We can see that during the hot photospheric phase, the spectra show a similar optical continuum shape to the average SLSN, though the lines in SN\,2023taz appear much narrower. This is partially due to artificial line broadening in the average spectrum from averaging many events with different velocities. Comparisons to individual objects such as iPTF13ajg \citep{Vreeswijk2014}, SN\,2017egm \citep{Xiang2017, Dong2017, Bose2018, Nicholl2017c, Zhu2023, Lin2023} and Gaia16apd \citep{Nicholl2017d, Yan2017b, Kangas2017} also show that this event looks overall similar to a typical SLSN but confirm the lower velocities. These comparison spectra were selected because their optical continua have similar temperatures and slopes to SN\,2023taz.

By 30 days post peak, SN\,2023taz has transitioned to the cooler photospheric stage with more prominent emission lines and P-Cygni profiles, as shown in the lower panel of Figure \ref{fig:spectra}. At these latter phases we can make a more direct comparison to the average spectra from \citet{Aamer2025}. We see many similarities in both the overall shape of the underlying continuum and in the specific lines. There are visible P-Cygni profiles from a blend of Fe II around $\sim$3600\,\AA, Ca II $\lambda\lambda$3934,3968, and a weak feature $\sim6300$\,\AA\ possibly due to [O I] $\lambda$6355. The feature around 4500\,\AA\ is likely due to a blend of Fe II with possible contributions from Mg I] $\lambda$4571. However, we note there is weaker Fe emission around 5500\,\AA\ compared to the population average. 

Another difference from the averages is the asymmetrical profile of the feature $\sim$4500 in the later spectra. The line is steeper on the red side, inconsistent with a typical P-Cygni profile. There is also a bump on the blue side of the feature, though this is coincident with a known Fe II line at 4447\,\AA. However, the question remains as to why this asymmetrical profile is visible in SN\,2023taz but not in the average spectrum. The asymmetry could be due to a lower velocity iron component, which would result in a narrower profile and less blending. To check this, we measured the velocities of the nearby blend of Fe II lines at 4924\,\AA, 5018\,\AA, and 5169\,\AA\ in the spectra of SN\,2023taz. We used the Markov chain Monte Carlo (MCMC) template matching code written by \citet{Modjaz2016}. The resulting velocities and phases are shown in Table \ref{tab:fe_lines}. The method did not produce a reliable fit for the low signal spectrum at +34.8 days, and so this measurement has been excluded. Comparing the velocities in Table \ref{tab:fe_lines} to the average values from \citet{Aamer2025}, we see that the velocity is on the lower end of the population but still comparable to other objects.  However, we include the caveat that the spectra before a phase of 10 days may be influenced by contamination from Fe III \citep{Liu2017}. Regardless, it appears that this unusual line profile cannot easily be explained by low velocities. Instead, the skewed profile likely reflects intrinsic asymmetries in the ejecta. There is tentative support for this interpretation from the Ca II $\lambda\lambda$3934, 3968 feature, where a similar asymmetry might be present. However, due to the low SNR in this region, we cannot make a definitive statement. There may also be support for this interpretation in the spectrum obtained at +34.8 days. Although much noisier, there may be a separate resolved bump on the blue side of both of these profiles. Nevertheless, these asymmetric features point towards ejecta asymmetry as the most likely source of the observed line profile deviations.

\begin{table}
	\centering
	\caption[Iron Line Velocities]{Fe II $\lambda$5169 line velocities}
	\label{tab:fe_lines}
	\begin{tabular}{cc}
		\hline \hline
            Phase (days) & Velocity (km\,s$^{-1}$)\\
            \hline
            8.0 & $13\,600 \pm 2800$ \\
            9.3 & $10\,200 \pm 2800$ \\
            39.2 & $9100 \pm 1200$ \\
            59.7 & $7800 \pm 800$ \\

		\hline \hline
	\end{tabular}
\end{table}

\section{Investigation of the UV Deficit}\label{sec:UV}

\begin{figure}
    \centering
    \includegraphics[width=\columnwidth]{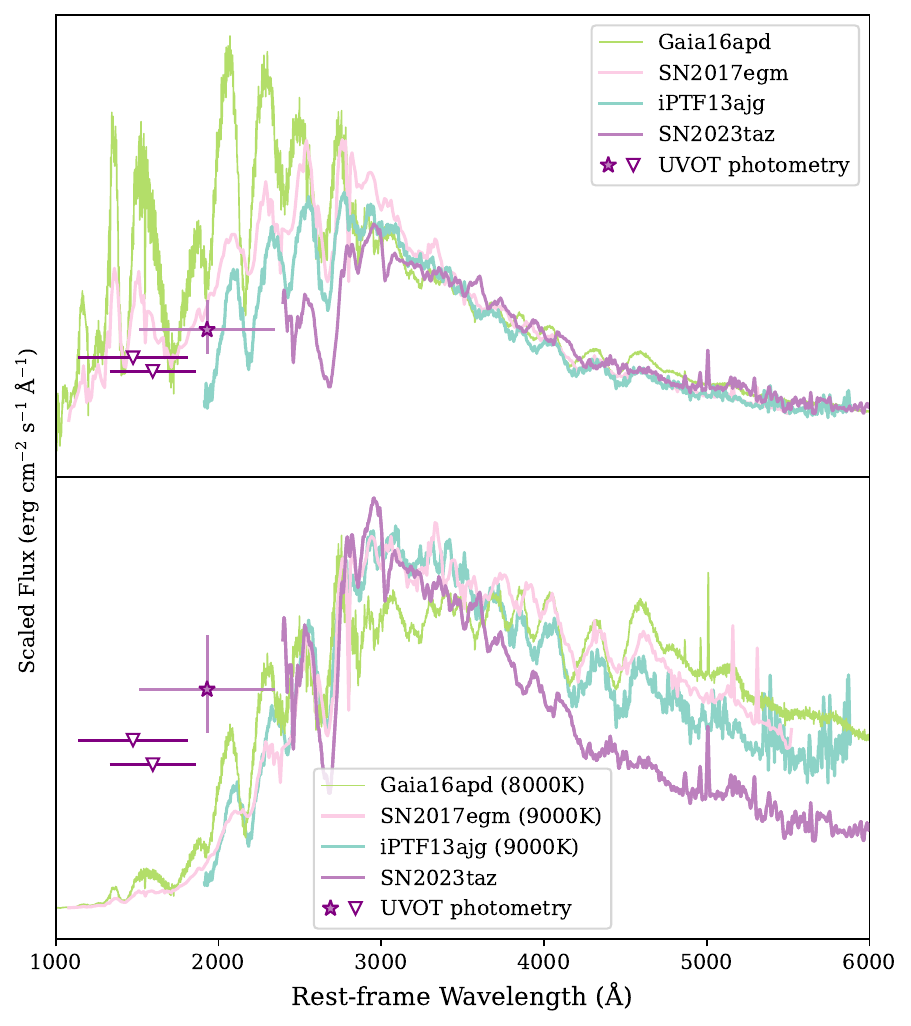}
    \caption{UV spectral comparisons for SN\,2023taz, Gaia16apd, SN\,2017egm, and iPTF13ajg. The UVOT photometry for SN\,2023taz are the measurements taken closest to peak. Photometric upper limits are indicated by inverted triangles. \textit{Top:} Observed optical and UV spectra around maximum light. The optical sections of the spectra are comparable between the three events, but they diverge in the UV, particularly in the depth of the Mg II $\lambda\lambda$2798, 2803 feature. \textit{Bottom:} The spectra for Gaia16apd, SN\,2017egm, and iPTF13ajg have been warped to blackbody temperatures of 8000\,K, 9000\,K, and 9000\,K, respectively. These temperatures were chosen to match the continuum shape around the Mg II UV absorption line to that observed in SN\,2023taz. However, after this warping to match the NUV, neither the optical continuum shapes nor the FUV fluxes are consistent with SN\,2023taz, showing that temperatures difference alone cannot explain the differences in the NUV colours. Moreover, after matching the UV continuum levels it is apparent that the equivalent width of the Mg II absorption line is significantly greater in SN\,2023taz.
    \label{fig:UV}}
\end{figure}

We have established SN\,2023taz to be a SLSN that is very luminous but otherwise quite typical in the optical. Its most surprising feature is the very red UV-optical colour. We now discuss possible physical origins.

Figure \ref{fig:colour} indicated a UV flux deficit at early times, using the UV-optical colours. We confirm this in the top panel of Figure \ref{fig:UV} using the early spectra of SN\,2023taz and comparison events with UV data from Figure \ref{fig:spectra}. Although SN\,2023taz, Gaia16apd, SN\,2017egm, and iPTF13ajg all show quite similar optical spectra, the fluxes diverge in the UV.


\subsection{Extinction} \label{sec:extinction_UV}

One possible cause of a redder UV–optical SED is extinction. Extinction curves of the Milky Way include a prominent bump thought to be due to the presence of graphite or polycyclic aromatic hydrocarbons \citep{Draine2003}. This broad feature spans over 1000\,\AA\ centred at 2175\,\AA\, and therefore would effect the observed flux in a wavelength range corresponding roughly  to the UVOT $UVM2$ filter.
However, at $z=0.407$, an observer frame wavelength of 2175\,\AA\ corresponds to only $\sim$1500\,\AA\ in the SN rest-frame, so cannot explain the suppression we see at longer rest-frame wavelengths of $\sim2000-3000$\,\AA\ (Figure \ref{fig:UV}).

Although data have been corrected for foreground extinction, host galaxy extinction was neglected in our SED analysis in Section \ref{sec:lc}. Analysis of the host galaxy spectrum (Section \ref{sec:host}) suggested that this was negligible, but to further confirm this we also check for Na I D absorption lines within the spectrum, which are often used a tracer for line-of-sight dust extinction \citep{Poznanski2012}. We find no evidence for Na I absorption within the series of spectra. 
The high temperatures needed to match the optical-only colour temperature for SN\,2023taz also indicate that extinction is unlikely to be the mechanism for the UV suppression, as this would work to redden the whole spectrum. 

Taken together, the lack of Na I absorption, the faint host, the \textsc{MOSFiT} modelling, the colour temperature evolution, and the Balmer decrements all lead to the same conclusion: extinction from the host galaxy or local environment is negligible, and dust extinction cannot be the primary cause of the UV suppression in SN\,2023taz.

\subsection{Temperature} \label{sec:temperature_UV}

One explanation for the lower UV flux levels could be due to a lower temperature (though as discussed in section \ref{sec:BB_params} the temperature measured from the optical data appears fairly typical). To investigate this in more detail, we scaled the spectra of Gaia16apd, SN\,2017egm, and iPTF13ajg to match the UV continuum shape on either side of the Mg II line in SN\,2023taz. This is done by dividing the spectra by the best fit blackbody temperature ($\sim$17\,000\,K, $\sim$14\,000\,K, and $\sim$15\,000\,K respectively), and multiplying by a new blackbody with a lower temperature following the methodology in \citet{Nicholl2017d}. We show the results of these transformations in the bottom panel of Figure \ref{fig:UV}. This allows us to test how much cooler these SLSNe would need to be in order to reproduce the near-UV colours of SN\,2023taz. Achieving a match to the UV spectrum between 2500-3000\AA\ in SN\,2023taz required scaling Gaia16apd to a temperature of 8000\,K, and SN\,2017egm and iPTF13ajg both to a temperature of 9000\,K. Although the continuum shape can be reproduced by the lower temperature, the depth of the Mg II line is difficult to reproduce. The warped spectra also have a much lower flux at $\sim$2000\AA\ than SN\,2023taz. Most importantly, scaling to a cooler continuum also vastly alters the optical SEDs, which no longer resemble SN\,2023taz. The spectra are also no longer physically self-consistent, as the dominant O II lines would not form at these lower temperatures. From this we can conclude that the colours of SN\,2023taz cannot be self-consistently reproduced by simply warping a typical SLSN spectrum to a cooler temperature.

\subsection{Absorption} \label{sec:absorption_UV}

%
Having ruled out extinction or a cooler photosphere as the cause of the unusually red colours of SN\,2023taz, we are left to consider enhanced UV line absorption as an explanation for the observed UV deficit. 

Another object that showed a flux deficit in the UV compared to most SLSNe is SN\,2017dwh. However, as shown in Figure \ref{fig:colour}, this event also had a much redder $g-r$ colour than SN\,2023taz. The spectra in Figure \ref{fig:spectra} show that the reason for this is enhanced absorption from Fe-group elements which can be seen with the broad absorption trough from Co II at $\sim$3200\,\AA\ in the spectrum at +8.31 days \citep{Blanchard2019}. This transitions to more of an overall line blanketing in the spectrum at +41.05 days which can be seen by the extremely red spectral shape. However, in the case of SN\,2023taz we do not see evidence for a large amount of Fe-group elements both from the lack of Fe II emission $\sim$5500\,\AA, but also the lack of absorption seen from this Co II line. Instead another explanation is needed to explain this level of UV absorption.

The warping in Figure \ref{fig:UV} indicated that enhanced absorption by intermediate mass elements such as Mg could account for some of the UV deficit. To quantify the strength of the UV line absorption, we measured the pseudo–equivalent width (pEW) of the broad absorption feature centred around 2670\,\AA, attributed primarily to Mg II with possible contributions from C II. We estimated the continuum as a straight line between the blue and red peaks of the feature and integrated the fractional depth. We find EW(Mg II) $\approx 75$\,\AA\ for SN\,2023taz, compared to 47\,\AA, 51\,\AA, and 44\,\AA\ for Gaia16apd, SN\,2017egm, and iPTF13ajg respectively. In SN\,2023taz, this absorption feature is significantly stronger, appearing $\sim$50\% more pronounced. 

Assuming the line is not saturated, the EW scales with the column density of the absorbing element. The implication is that SN\,2023taz has a larger quantity of intermediate-mass elements, particularly Mg, above the photosphere at this epoch. This could be the result of several factors. In typical SLSN models, Mg is produced in the ashes of explosive carbon burning. If the progenitor was highly stripped, with most of the He and some of the C layers removed, this could allow Mg-rich layers to be visible closer to the surface. However, the deficit appearing at shorter UV wavelengths (around 2400\AA) where strong carbon lines are expected \citep{Vreeswijk2014, Yan2017b}, makes this scenario less likely, as it would require both more Mg and C.

Alternatively, the UV suppression could reflect enhanced mixing of Mg into the outer ejecta. One possible way to increase mixing is a powerful central engine that drives fluid instabilities \cite{Chen2016, Chen2020, Suzuki2021}. This would be consistent with the high luminosity of SN\,2023taz, though it raises the question of why a similar effect is not seen in Gaia16apd, a similarly luminous (if shorter lived) event. 

A third possibility is that the photosphere in SN\,2023taz has receded further by peak light, into a deeper Mg-rich zone. This is consistent with the low expansion velocities measured from O II lines, but perhaps in tension with the high luminosity, which would typically sustain a more extended photosphere.

We cannot distinguish definitively between these scenarios, but all suggest that the red UV–optical colour in SN\,2023taz is best explained by unusually strong UV line blanketing rather than differences in temperature or extinction. Obtaining spectroscopy of future SLSNe further into the UV will help to illuminate the reason.

\section{Conclusions}\label{sec:conclusion}

The analysis in this paper focussed on the SLSN SN\,2023taz, presenting the optical and UV photometry and spectra obtained across a period of over 300 days. This event is one of the brightest SLSNe to date with an absolute magnitude of $M_{g,\rm{peak}}=-22.75 \pm 0.03$\,mag, and an estimated energy radiated of $E=2.9 \times 10^{51}$\,erg. Light curve modelling indicates that parameters required in the central engine theory are not individually noteworthy, but when combined place SN\,2023taz in a unusual region of the parameter space with $P_{\rm{spin}} = 2.25^{+0.33}_{-0.36}$\,ms, $\log(B_{\perp}/\rm{G}) = 13.73^{+0.13}_{-0.14}$, and $M_{\rm{ej}}=2.57^{+0.66}_{-0.48}$\,\msun. In the magnetar framework, SLSNe achieve high luminosities when the spin-down timescale of the central engine and the photon diffusion timescale of the ejecta are well matched, or when the magnetar is born spinning near its maximal rotation rate. In the case of SN\,2023taz, both of these conditions are met, placing it near the upper end of the SLSN luminosity distribution.

SED fits to the multi-colour light curve indicate high temperatures reaching a peak of $16\,600\pm4000$\,K just after peak. Despite this high peak temperature, the UV colour of this event is the reddest observed for a SLSNe with a colour $UVW1-r=1.83\pm 0.23$ compared to the population average at this phase of $UVW1-r=-0.73^{+0.39}_{-0.26}$. This is despite a relatively normal $g-r$ colour evolution compared to the population.

Taking SLSNe with well-observed UV spectra and warping these to match the UV spectral shape of SN\,2023taz requires temperatures of $\sim8000-9000$\,K, inconsistent with the shapes of their optical SEDs. This shows that a lower temperature than other SLSNe cannot explain the UV deficit.
Moreover, there is no evidence for significant extinction in SN\,2023taz, and comparison to SN\,2017dwh showed that iron line blanketing does not produce the observed spectral shape. The equivalent width of the Mg II absorption feature around 2800\,\AA\ in SN\,2023taz is measured to be 75\,\AA, significantly larger than in Gaia16apd, SN\,2017egm, and iPTF13ajg, which have equivalent widths of $\lesssim50$\,\AA, suggesting that the UV deficit is most likely due to a higher column density of Mg and potentially other intermediate-mass elements.

Future surveys like LSST will find large numbers of SLSNe, but the increase in detection rate comes from the ability of deeper surveys to detect more distant events. LSST will detect $\sim10\,000$ SLSNe per year out to $z\sim3$ \citep{Villar2018}. However, beyond $z\gtrsim1$, optical observations will probe only rest-frame UV emission, and beyond $z\gtrsim4$ near-IR observations will probe the rest-frame UV. Thus photometric identification and spectroscopic classification will be required based on this rest-frame UV \citep{Barbary2009, Cooke2012, Pan2017, Smith2018, Curtin2019}. The case of SN\,2023taz highlights that SLSNe may exhibit greater diversity in their UV properties than previously appreciated, even when appearing typical in the optical. This underlines the critical need for joint UV-optical observations of low redshift SLSNe. Understanding how UV diversity correlates with optical features will be essential to reliably identify and characterise these events as we move into an era of high-redshift transient surveys.

\begin{acknowledgments}
AA, MN and CA are supported by the European Research Council (ERC) under the European Union’s Horizon 2020 research and innovation programme (grant agreement No.~948381) and by UKSA/STFC Grant No. ST/Y000692/1.

JC and NVB acknowledge funding from the Australian Research Council Discovery Project DP200102102 and the Australian Research Council Centre of Excellence for Gravitational Wave Discovery (OzGrav), CE170100004 and CE230100016.

FP acknowledges support from the Spanish Ministerio de Ciencia, Innovación y Universidades (MICINN) under grant numbers PID2022-141915NB-C21.

T-WC acknowledges the financial support from the Yushan Fellow Program by the Ministry of Education, Taiwan (MOE-111-YSFMS-0008-001-P1) and the National Science and Technology Council, Taiwan (NSTC grant 114-2112-M-008-021-MY3).

CPG acknowledges financial support from the Secretary of Universities and
Research (Government of Catalonia) and by the Horizon 2020 Research and
Innovation Programme of the European Union under the Marie Sk\l{}odowska-Curie and the Beatriu de Pin\'os 2021 BP 00168 programme, from the
Spanish Ministerio de Ciencia e Innovaci\'on (MCIN) and the Agencia
Estatal de Investigaci\'on (AEI) 10.13039/501100011033 under the
PID2023-151307NB-I00 SNNEXT project, from Centro Superior de Investigaciones Cient\'ificas (CSIC) under the PIE project 20215AT016 and the program Unidad de Excelencia Mar\'ia de Maeztu CEX2020-001058-M, and from the Departament de Recerca i Universitats de la Generalitat de Catalunya through the 2021-SGR-01270 grant.

RKT acknowledges support by the NKFIH/OTKA FK-134432 grant of the National Research, Development and Innovation (NRDI) Office of Hungary.

TEMB is funded by Horizon Europe ERC grant no. 101125877.

BW is supported by UKRI's STFC studentship grant funding, project reference ST/X508871/1.

Based on observations collected at the European Organisation for Astronomical Research in the Southern Hemisphere, Chile, under ESO programme 114.27AG.001, and as part of ePESSTO+ (the advanced Public ESO Spectroscopic Survey for Transient Objects Survey – PI: Inserra), under ESO program IDs 112.25JQ. 

This work has made use of data from the Asteroid Terrestrial-impact Last Alert System (ATLAS) project. ATLAS is primarily funded to search for near earth asteroids through NASA grants NN12AR55G, 80NSSC18K0284, and 80NSSC18K1575; byproducts of the NEO search include images and catalogs from the survey area. The ATLAS science products have been made possible through the contributions of the University of Hawaii Institute for Astronomy, the Queen's University Belfast, the Space Telescope Science Institute, and the South African Astronomical Observatory.

The Pan-STARRS1 Surveys (PS1) have been made possible through contributions of the Institute for Astronomy, the University of Hawaii, the Pan-STARRS Project Office, the Max-Planck Society and its participating institutes, the Max Planck Institute for Astronomy, Heidelberg and the Max Planck Institute for Extraterrestrial Physics, Garching, The Johns Hopkins University, Durham University, the University of Edinburgh, Queen's University Belfast, the Harvard-Smithsonian Center for Astrophysics, the Las Cumbres Observatory Global Telescope Network Incorporated, the National Central University of Taiwan, the Space Telescope Science Institute, the National Aeronautics and Space Administration under Grant No. NNX08AR22G issued through the Planetary Science Division of the NASA Science Mission Directorate, the National Science Foundation under Grant No. AST-1238877, the University of Maryland, and Eotvos Lorand University (ELTE).

Based on observations made with the Nordic Optical Telescope (NOT), owned in collaboration by the University of Turku and Aarhus University, and operated jointly by Aarhus University, the University of Turku and the University of Oslo, representing Denmark, Finland and Norway, the University of Iceland and Stockholm University at the Observatorio del Roque de los Muchachos, La Palma, Spain, of the Instituto de Astrofisica de Canarias. The data presented here were obtained with ALFOSC, which is provided by the Instituto de Astrofisica de Andalucia (IAA) under a joint agreement with the University of Copenhagen and NOT.

This research made use of the ``K-corrections calculator'' service available at http://kcor.sai.msu.ru/.

\end{acknowledgments}

\bibliography{lib}{}
\bibliographystyle{aasjournalv7}



\end{document}